\documentclass[aps,preprint]{revtex4}%
\usepackage{amsfonts}
\usepackage{amsmath}
\usepackage{amssymb}
\usepackage{graphicx}%
\begin{document}
\title[Relativity Controversies]{Examples and Comments Related to Relativity Controversies}
\author{Timothy H. Boyer}
\affiliation{Department of Physics, City College of the City University of New York, New
York, New York 10031}

\begin{abstract}
Recently Mansuripur has called into question the validity of the Lorentz force
in connection with relativitistic electromagnetic theory. \ Here we present
some very simple point-charge systems treated through order $v^{2}/c^{2}$ in
order to clarify some aspects of relativistic controversies both old and new.
\ In connection with the examples, we confirm the validity of the relativistic
conservation laws. \ The relativistic examples make clear that external forces
may produce a vanishing torque in one inertial frame and yet produce a
non-zero torque in another inertial frame, and that the conservation of
angular momentum will hold in both frames. \ We also discuss a relativistic
point-charge model for a magnetic moment and comment on the interaction of a
point charge and a magnetic moment. \ Mansuripur's claims of incompatibiltiy
between the Lorentz force and relativity are seen to be invalid.

\end{abstract}
\maketitle

\section{Introduction}

\subsection{Mansuripur's Controversial Claim}

The Lorentz force on point charges, Maxwell's equations for the
electromagnetic fields, and appropriate boundary conditions on the
differential equations- these three aspects form the foundations of
relativistic classical electron theory appearing in textbooks of classical
electrodynamics.\cite{Griffiths}\cite{Jackson} \ Therefore it seems surprising
when an article by Mansuripur, published in the prominent physics research
journal Physical Review Letters, claims to present "incontrovertible
theoretical evidence of the incompatibility of the Lorentz law with the
fundamental tenets of special relativity."\cite{Mansuripur} \ The claim of
incompatibility arises in connection with the interaction of a point charge
and a magnetic moment and is based on the statement, "The appearance of this
torque in the xyz frame in the absence of a corresponding torque in the x'y'z'
frame is sufficient proof of the inadequacy of the Lorentz
law."\cite{Mansuripur} \ Mansuripur's claim is highlighted by an article in
Science under the title "Textbook Electrodynamics May Contradict
Relativity."\cite{Science} \ The surprising claim of incompatibility arises in
connection with 1) torques and involving 2) the interaction of a point charge
and a magnetic moment. \ Now these two aspects are familiar in long-lasting
controversies related to relativity. The issue of torques, which appeared in
some inertial frames but not others, formed the basis for the experiments of
Trouton and Noble\cite{TN} back in 1903 and is still discussed as a relativity
paradox.\cite{TNarticles} \ The interaction of a magnetic moment with an
electric field (such as provided by a point charge) is at the heart of the
controversies involving "hidden momentum,"\cite{Shockley} the Aharonov-Bohm
effect,\cite{AB} and the Aharonov-Casher effect.\cite{AC} \ Indeed, recently
Griffiths provided a review of some of the clashing points of view in his
resource letter on electromagnetic momentum.\cite{RLetter} \ 

In this article we point out that Mansuripur's bold claim of relativistic
incompatibility for the Lorentz force is invalid for two basic reasons.
\ First, it is a fact of relativity that the absence of a torque in one
inertial frame does not preclude the existence of a torque due to the same
forces as seen in a second inertial frame. \ Second, the interaction of a
point charge and a magnetic moment has never been described at the level of
detail required for a relativistic understanding, and therefore this
interaction cannot be used as a basis for a claim of inconsistency.
\ Specifically, there has never been a discussion of even the nonrelativistic
\textit{internal} electromagnetic stresses required for the stability of a
magnetic moment in the presence of an electric field. \ This same lack of
necessary relativistic detail is present in many of the discussions listed in
Griffiths' resource letter.\cite{RLetter}

In fairness to Mansuripur, we should note that his real interest is in the
relativistic description of polarized and magnetized materials; his
incompatibility claim arises from attempts to understand this complicated
problem involving materials. \ In the present article, we have a far more
modest aim. \ We will deal only with very simple point charge systems which
can be understood in relativistic detail. \ We present three simple
charged-particle systems which are relativistic (to order $v^{2}/c^{2})$ and
which are related to past and present relativistic controversies. \ These
systems highlight some basic aspects of relativity and remind us that
Mansuripur's claim of incompatibility can not be taken seriously.

\subsection{Outline of the Presentation}

Before introducing the point-charge examples, we deal with some relativistic
preliminaries. \ We remind the reader of the relativistic conservation laws
and of the relativistic force transformation laws which will be used in the
examples. \ Our analysis makes use of the Darwin Lagrangian which involves
interacting charged point masses and is known to represent the relativistic
interactions of classical electrodynamics through order $v^{2}/c^{2}.$ \ We
review the expressions for energy, momentum, angular momentum, and the
equations of motion which follow from the Darwin Lagrangian.

Our first example involves two point charges held at rest by external forces.
\ This system is related to the old "4/3-problem" for the classical model of
the electron,\cite{43prob} where stabilizing forces must be present to
maintain the electrostatic configuration. \ The point charge system is also
related to the Trouton-Noble experiment and to Mansuripur's implied assertion
that the absence of a torque in one inertial frame means its absence in all
inertial frames. \ We note that when external forces are present,
electromagnetic energy and momentum do not transform as a Lorentz 4-vector.
\ To obtain a physical appreciation of this situation, we consider two charged
point particles of equal mass $m$ and charge $e$ which are approaching each
other symmetrically in their center-of-energy inertial frame. \ When the
charges come instantaneously to rest in this center-of-energy frame, external
forces are applied. \ However, the application of the external forces which is
simultaneous in one inertial frame need not be simultaneous in a relatively
moving frame; furthermore, the external forces need not be along a common line
of action. \ Thus we find that in the moving frame the external forces may
introduce energy, linear momentum, and angular momentum.

Second we consider a model for a \ magnetic moment which consists of two point
particles of charges $e$ and $-e$ and masses $m$ and $M$ which are in uniform
circular motion in on inertial frame. \ We find the associated system magnetic
moment. \ We consider the aspects of the magnetic moment as seen in a second,
relatively moving inertial frame and note the appearance of a non-zero
electric dipole moment in the frame where the center of energy of the magnetic
dipole is moving.. \ 

Thirdly, we consider the introduction of an additional point charge $q$ into
the inertial frame with the magnetic moment model. \ We point out that the
magnetic moment model is no longer stable, not even to zero order in $v/c$.
\ Any description of the interaction of the magnetic moment and the point
charge must account explicitly for the forces or accelerations which are
associated with the response of the magnetic moment model to the presence of
an external electric field across the magnetic moment. \ Although such an
account has been given for the case when \textit{external} forces are applied
to the charges of the magnetic moment model,\cite{b44} a convincing account of
the interaction has never be given for a system where relativistic
\textit{internal} stresses are assumed to exist. \ In the absence of such a
detailed relativistic description, Mansuripur's claims of an incompatibility
with relativity can not be taken seriously. \ 

\section{Relativistic Preliminaries}

\subsection{Relativistic Conservation Laws}

In the examples to follow, we will turn repeatedly to the relativistic
conservation laws. \ These include the familiar conservation laws for energy,
linear momentum, and angular momentum. \ However, the last (and only
specifically relativistic) conservation law involves the uniform motion of the
center of energy. \ In the presence of an external force $\mathbf{F}%
_{ext,\,i}$ on particle $i$ at position $\mathbf{r}_{i}$ with velocity
$\mathbf{v}_{i}$, the conservation laws for a Lorentz-invariant mechanical
system or field theory take the following forms.\cite{CofE}\cite{Implications}
The sum of the external forces on the system gives the time rate of change of
the system linear momentum $\mathbf{P}$
\begin{equation}
\sum_{i}\mathbf{F}_{ext,\,i}=\frac{d\mathbf{P}}{dt}.\label{1}%
\end{equation}
The total power delivered to the system by the external forces gives the time
rate of change of the system energy $U$
\begin{equation}
\sum_{i}\mathbf{F}_{ext,\,i}\cdot\mathbf{v}_{i}=\frac{dU}{dt}.\label{2}%
\end{equation}
The sum of the external torques gives the time rate of change of system
angular momentum $\mathbf{L}$
\begin{equation}
\sum_{i}\mathbf{r}_{i}\times\mathbf{F}_{ext,\,i}=\frac{d\mathbf{L}}%
{dt}.\label{3}%
\end{equation}
The external power-weighted position equals the time rate of change of (the
total energy $U$ times the center of energy $\overrightarrow{\mathcal{X}})$
minus $c^{2}$ times the system momentum:
\begin{equation}
\sum_{i}(\mathbf{F}_{ext,\,i}\cdot\mathbf{v}_{i})\mathbf{r}_{i}=\frac{d}%
{dt}(U\overrightarrow{\mathcal{X}})-c^{2}\mathbf{P}.\label{4}%
\end{equation}
The center of energy $\overrightarrow{\mathcal{X}}$ is defined so that
\begin{equation}
U\overrightarrow{\mathcal{X}}=\sum_{i}U_{i}\mathbf{r}_{i}+\!\int
\!d^{3}r\,u(\mathbf{r})\mathbf{r},\label{5}%
\end{equation}
where $U_{i}$ is the mechanical energy (rest energy plus kinetic energy) of
the $i$th particle and $u(\mathbf{r})$ is the continuous system energy density
at position $\mathbf{r}$. This last conservation law in Eq. (\ref{4})
expresses the continuous flow of energy in Lorentz-invariant systems. In an
isolated system where no external forces are present, the linear momentum
$\mathbf{P}$, energy $U$, and angular momentum $\mathbf{L}$ are all constants
in time, and the center of energy $\overrightarrow{\mathcal{X}}$ moves with
constant velocity $d\overrightarrow{\mathcal{X}}/dt=c^{2}\mathbf{P}/U$,
because the energy $U$ and momentum $\mathbf{P}$ in Eq.~(\ref{4}) are both constant.

\subsection{Lorentz Transformation of Forces}

Although the 4-vector Lorentz transformations for spacetime displacements and
for energy-momentum are familiar to students, the force transformations are
usually less familiar. \ The Lorentz transformation of a force $\mathbf{F}%
=\widehat{i}F_{x}+\widehat{j}F_{y}+\widehat{k}F_{z}$ acting on a point
particle moving with velocity $\mathbf{u}$ in the $S$ inertial frame is given
in the $S^{\prime}$ inertial frame (moving with velocity $\mathbf{V}%
=\widehat{i}V$ along the $x$-axis of the $S$ frame) by%
\begin{equation}
F_{x}^{\prime}=\frac{F_{x}-\mathbf{u}\cdot\mathbf{F}V/c^{2}}{1-u_{x}V/c^{2}%
}\label{F1}%
\end{equation}%
\begin{equation}
F_{y}=\frac{F_{y}}{\gamma(1-u_{x}V/c^{2})},\text{ \ \ \ }F_{z}=\frac{F_{z}%
}{\gamma(1-u_{x}V/c^{2})}\label{F2}%
\end{equation}
These transformation formulae follow from use of the Lorentz force and the
Lorentz transformations for the electric and magnetic fields.

\subsection{The Darwin Lagrangian}

In the examples to follow, we will not discuss electromagnetic interactions to
all orders in $v/c,$ since such an analysis can become quite complicated.
\ Rather for simplicity sake, we will restrict our analysis to order
$v^{2}/c^{2}.$ \ The electromagnetic interaction of two point particles of
charges $e_{1}$, $e_{2}$, and masses $m_{1},m_{2}$ can be described through
order $v^{2}/c^{2}$ by the Lagrangian%
\begin{align}
\mathcal{L} &  =-m_{1}c^{2}+\frac{1}{2}m_{1}\left(  \mathbf{v}_{1}^{2}%
+\frac{1}{4}\frac{(\mathbf{v}_{1}^{2})^{2}}{c^{2}}\right)  -m_{2}c^{2}%
+\frac{1}{2}m_{2}\left(  \mathbf{v}_{2}^{2}+\frac{1}{4}\frac{(\mathbf{v}%
_{2}^{2})^{2}}{c^{2}}\right)  -\frac{e_{1}e_{2}}{|\mathbf{r}_{1}%
-\mathbf{r}_{2}|}\nonumber\\
&  +\frac{e_{1}e_{2}}{2c^{2}}\left[  \frac{\mathbf{v}_{1}\cdot\mathbf{v}_{2}%
}{|\mathbf{r}_{1}-\mathbf{r}_{2}|}+\frac{\mathbf{v}_{1}\cdot(\mathbf{r}%
_{1}-\mathbf{r}_{2})\mathbf{v}_{2}\cdot(\mathbf{r}_{1}-\mathbf{r}_{2}%
)}{|\mathbf{r}_{1}-\mathbf{r}_{2}|^{3}}\right]  \label{e1}%
\end{align}
first introduced by C. G. Darwin.\cite{Darwin} \ Here the terms involving the
masses arise from the relativistic \textit{mechanical} Lagrangian
\begin{equation}
\mathcal{L}_{mech}=-\frac{mc^{2}}{(1-v^{2}/c^{2})^{1/2}}\approx-mc^{2}%
+\frac{1}{2}m_{1}\left(  \mathbf{v}_{1}^{2}+\frac{1}{4}\frac{(\mathbf{v}%
_{1}^{2})^{2}}{c^{2}}\right)  \label{e1a}%
\end{equation}
and the terms involving the charges $e_{1},e_{2}$ correspond to
electromagnetic field contributions $-%
{\textstyle\int}
d^{3}r(\mathbf{E}^{2}-\mathbf{B}^{2})$ with the divergent self-energy
integrals omitted. \ The total momentum is the sum $\mathbf{P}=\mathbf{p}%
_{1}+\mathbf{p}_{2}$ where the canonical momentum $\mathbf{p}_{i}$ of the
$i$th particle is given by%
\begin{equation}
\frac{\partial\mathcal{L}}{\partial\mathbf{v}_{i}}=\mathbf{p}_{i}=m_{i}\left(
1+\frac{v_{i}^{2}}{2c^{2}}\right)  \mathbf{v}_{i}+\frac{e_{i}e_{j}}{2c^{2}%
}\left(  \frac{\mathbf{v}_{j}}{|\mathbf{r}_{i}-\mathbf{r}_{j}|}+\frac
{\mathbf{v}_{j}\cdot(\mathbf{r}_{i}-\mathbf{r}_{j})(\mathbf{r}_{i}%
-\mathbf{r}_{j})}{|\mathbf{r}_{i}-\mathbf{r}_{j}|^{3}}\right)  \label{e2}%
\end{equation}
which includes the $v^{2}/c^{2}$ contribution from the mechanical momentum of
the $i$th particle
\begin{equation}
\mathbf{p}_{mech\text{ }i}=\frac{m_{i}\mathbf{v}_{i}}{(1-v_{i}^{2}%
/c^{2})^{1/2}}\mathbf{\approx}m_{i}\left(  1+\frac{v_{i}^{2}}{2c^{2}}\right)
\mathbf{v}_{i}\label{e3}%
\end{equation}
and also electromagnetic field momentum associated with the electrostatic
field of the $i$th particle and the magnetic field of the $j$th particle. The
total angular momentum is the sum $\mathbf{L=L}_{1}+\mathbf{L}_{2}$ where the
canonical angular momentum $\mathbf{L}_{i}$ of the $i$th particle is given by
\begin{equation}
\mathbf{L}_{i}=\mathbf{r}_{i}\times\mathbf{p}_{i}=\mathbf{r}_{i}\times\left[
m_{i}(1+\frac{v_{i}^{2}}{2c^{2}})\mathbf{v}_{i}+\frac{e_{i}e_{j}}{2c^{2}%
}\left(  \frac{\mathbf{v}_{j}}{|\mathbf{r}_{i}-\mathbf{r}_{j}|}+\frac
{\mathbf{v}_{j}\cdot(\mathbf{r}_{i}-\mathbf{r}_{j})(\mathbf{r}_{i}%
-\mathbf{r}_{j})}{|\mathbf{r}_{i}-\mathbf{r}_{j}|^{3}}\right)  \right]
\label{e4}%
\end{equation}
The total system energy $U$ including the rest energy is given by%
\begin{align}
U &  =m_{1}c^{2}+\frac{1}{2}m_{1}\left(  \mathbf{v}_{1}^{2}+\frac{3}{4}%
\frac{(\mathbf{v}_{1}^{2})^{2}}{c^{2}}\right)  +m_{2}c^{2}+\frac{1}{2}%
m_{2}\left(  \mathbf{v}_{2}^{2}+\frac{3}{4}\frac{(\mathbf{v}_{2}^{2})^{2}%
}{c^{2}}\right)  \nonumber\\
&  +\frac{e_{1}e_{2}}{|\mathbf{r}_{1}-\mathbf{r}_{2}|}+\frac{e_{1}e_{2}%
}{2c^{2}}\left[  \frac{\mathbf{v}_{1}\cdot\mathbf{v}_{2}}{|\mathbf{r}%
_{1}-\mathbf{r}_{2}|}+\frac{\mathbf{v}_{1}\cdot(\mathbf{r}_{1}-\mathbf{r}%
_{2})\mathbf{v}_{2}\cdot(\mathbf{r}_{1}-\mathbf{r}_{2})}{|\mathbf{r}%
_{1}-\mathbf{r}_{2}|^{3}}\right]  \label{e5}%
\end{align}
and the center of energy $\overrightarrow{\mathcal{X}}$ is given by
\begin{align}
U\overrightarrow{\mathcal{X}} &  =\left[  m_{1}c^{2}+\frac{1}{2}m_{1}\left(
\mathbf{v}_{1}^{2}+\frac{3}{4}\frac{(\mathbf{v}_{1}^{2})^{2}}{c^{2}}\right)
\right]  \mathbf{r}_{1}+\left[  m_{2}c^{2}+\frac{1}{2}m_{2}\left(
\mathbf{v}_{2}^{2}+\frac{3}{4}\frac{(\mathbf{v}_{2}^{2})^{2}}{c^{2}}\right)
\right]  \mathbf{r}_{2}\nonumber\\
&  +\left\{  \frac{e_{1}e_{2}}{|\mathbf{r}_{1}-\mathbf{r}_{2}|}+\frac
{e_{1}e_{2}}{2c^{2}}\left[  \frac{\mathbf{v}_{1}\cdot\mathbf{v}_{2}%
}{|\mathbf{r}_{1}-\mathbf{r}_{2}|}+\frac{\mathbf{v}_{1}\cdot(\mathbf{r}%
_{1}-\mathbf{r}_{2})\mathbf{v}_{2}\cdot(\mathbf{r}_{1}-\mathbf{r}_{2}%
)}{|\mathbf{r}_{1}-\mathbf{r}_{2}|^{3}}\right]  \right\}  \frac{(\mathbf{r}%
_{1}+\mathbf{r}_{2})}{2}\label{e6}%
\end{align}
Thus there are both mechanical and field contributions to the energy, linear
momentum, angular momentum, and center of energy. \ The equations of motion
for the $i$th particle follow from the Lagrangian as%
\begin{equation}
\frac{d\mathbf{p}_{i}}{dt}=\frac{\partial\mathcal{L}}{\partial\mathbf{r}_{i}%
}\label{e7}%
\end{equation}
where the vector character of the equation is understood to mean separate
treatment for each rectangular component. \ However, these equations of motion
can be rewritten in terms of the Lorentz force on the $i$th particle due to
the electric and magnetic fields of the other particle (the $j$th particle)%
\begin{equation}
\frac{d}{dt}\left[  \frac{m_{i}\mathbf{v}_{i}}{(1-v_{i}^{2}/c^{2})^{1/2}%
}\right]  \approx\frac{d}{dt}\left[  m_{i}\left(  1+\frac{v_{i}^{2}}{2c^{2}%
}\right)  \mathbf{v}_{i}\right]  =e_{i}\mathbf{E}_{j}\mathbf{(r}_{i}%
,t)+e_{i}\frac{\mathbf{v}_{i}}{c}\times\mathbf{B}_{j}(\mathbf{r}%
_{i},t)\label{e8}%
\end{equation}
where the electric and magnetic fields due to the $j$th particle are given
through order $v^{2}/c^{2}$ by\cite{assign}%
\begin{align}
\mathbf{E}_{j}(\mathbf{r}_{,}t) &  =e_{j}\frac{(\mathbf{r}-\mathbf{r}_{j}%
)}{|\mathbf{r}-\mathbf{r}_{j}|^{3}}\left[  1+\frac{1}{2}\frac{v_{j}^{2}}%
{c^{2}}-\frac{3}{2}\left(  \frac{\mathbf{v}_{j}\cdot(\mathbf{r}-\mathbf{r}%
_{j})}{c|\mathbf{r}-\mathbf{r}_{j}|}\right)  ^{2}\right]  \nonumber\\
&  -\frac{e_{j}}{2c^{2}}\left(  \frac{\mathbf{a}_{j}}{|\mathbf{r}%
-\mathbf{r}_{j}|}+\frac{\mathbf{a}_{j}\cdot(\mathbf{r}-\mathbf{r}%
_{j})(\mathbf{r}-\mathbf{r}_{j})}{|\mathbf{r}-\mathbf{r}_{j}|^{3}}\right)
\label{e9}%
\end{align}%
\begin{equation}
\mathbf{B}_{j}(\mathbf{r},t)=e_{j}\frac{\mathbf{v}_{j}}{c}\times
\frac{(\mathbf{r}-\mathbf{r}_{j})}{|\mathbf{r}-\mathbf{r}_{j}|^{3}}\label{e10}%
\end{equation}
We note that, in general, accelerations appear on both the left- and
right-hand sides of Eq. (\ref{e8}) since the electric field $\mathbf{E}%
_{j}(\mathbf{r}_{i},t)$ at the position $\mathbf{r}_{i\text{ }}$of the $i$th
particle due to the charge $e_{j}$ of the $j$th particle depends upon the
acceleration of the $j$th particle as in Eq. (\ref{e9}).

\section{Two Particles Held at Rest by External Forces}

\subsection{Two Point Charges Held at Rest }

The first system which we consider is two point particles, both of charge $e$
and mass $m,$ which are held at rest at positions $\mathbf{r}_{1}$ and
$\mathbf{r}_{2}$ with separation $|\mathbf{r}_{1}-\mathbf{r}_{2}|=l$ in an
inertial frame $S$. \ The external forces $\mathbf{F}_{1}=-\mathbf{F}_{2}$
holding the particles at rest must balance the electrostatic repulsion between
the charges $F_{1}=e^{2}/l^{2}=F_{2}.$ \ In the frame $S$, the system has
total energy $U=2mc^{2}+e^{2}/l$ (including contributions from the rest mass
and the electrostatic energy), has total momentum $\mathbf{P}=0,$ and total
angular momentum $\mathbf{L}=0.$ \ We will assume that the particles are
placed symmetrically about the origin of coordinates so that the center of
energy $\overrightarrow{\mathcal{X}}$ is also zero, $U\mathcal{X}_{E}%
=mc^{2}\mathbf{r}_{1}+mc^{2}\mathbf{r}_{2}+(e^{2}/l)(\mathbf{r}_{1}%
+\mathbf{r}_{2})/2=0$ since $\mathbf{r}_{1}=-\mathbf{r}_{2}.$

\subsection{Lorentz Transformation of Energy and Momentum in the Presence of
External Forces}

\subsubsection{The Lorentz-4-Vector Expectation}

Many students expect energy and momentum to transform as a Lorentz 4-vector.
\ However, this expectation may not be true when external forces are
present.\cite{b1985a} \ If we consider an inertial frame $S^{\prime}$ moving
with velocity $\mathbf{V}=\widehat{i}V$ along the $x$-axis of the inertial
frame $S$, then naive 4-vector expectation would suggest that in $S^{\prime}$
the energy and momentum are
\begin{align}
U^{\prime}  & =\gamma_{V}U\approx\left(  1+\frac{V^{2}}{2c^{2}}+\frac{3V^{4}%
}{8c^{4}}\right)  \left(  2mc^{2}+\frac{e^{2}}{l}\right)  \nonumber\\
& =2mc^{2}+2m\left(  \frac{V^{2}}{2}+\frac{3V^{4}}{8c^{2}}\right)  +\left(
1+\frac{V^{2}}{2c^{2}}\right)  \frac{e^{2}}{l}\text{ \ \ \ expected 4-vector
form }\label{Uex}\\
\mathbf{P}^{\prime}  & =-\frac{\gamma_{V}\mathbf{V}U}{c^{2}}\approx-2m\left(
1+\frac{V^{2}}{2c^{2}}\right)  \mathbf{V}-\frac{e^{2}}{lc^{2}}\mathbf{V}\text{
\ \ \ \ expected 4-vector form}\label{Pex}%
\end{align}
since
\begin{equation}
\gamma_{V}=\frac{1}{(1-V^{2}/c^{2})^{1/2}}=1+\frac{V^{2}}{2c^{2}}+\frac
{3V^{4}}{8c^{4}}+...\label{G}%
\end{equation}
\ In the inertial frame $S^{\prime},$ the particles are moving with velocities
$-\widehat{i}V$ so that the \textit{mechanical} contributions to the energy
and momentum are indeed of the 4-vector form
\begin{align}
U_{mech}^{\prime} &  =2mc^{2}\gamma_{V}=\gamma_{V}U_{mech}\approx
2mc^{2}+2m\left(  \frac{V^{2}}{2}+\frac{3}{8}\frac{V^{4}}{c^{4}}\right)
\label{U0}\\
\mathbf{P}_{mech}^{\prime} &  =-2m\mathbf{V}\gamma_{V}=-\gamma_{V}%
\mathbf{V}U_{mech}/c^{2}\approx-2m\left(  1+\frac{V^{2}}{2c^{2}}\right)
\mathbf{V}\label{P0}%
\end{align}
\ However, the \textit{electromagnetic field} contributions depend upon the
relative orientation between the velocity $\mathbf{V}$ and the particle
separation $\mathbf{r}_{1}-\mathbf{r}_{2}.$ \ From Eqs. (\ref{e5}) and
(\ref{e2}), we have%

\begin{align}
U_{em} &  =\frac{e_{1}e_{2}}{|\mathbf{r}_{1}-\mathbf{r}_{2}|}+\frac{e_{1}%
e_{2}}{2c^{2}}\left[  \frac{\mathbf{v}_{1}\cdot\mathbf{v}_{2}}{|\mathbf{r}%
_{1}-\mathbf{r}_{2}|}+\frac{\mathbf{v}_{1}\cdot(\mathbf{r}_{1}-\mathbf{r}%
_{2})\mathbf{v}_{2}\cdot(\mathbf{r}_{1}-\mathbf{r}_{2})}{|\mathbf{r}%
_{1}-\mathbf{r}_{2}|^{3}}\right]  \label{U1}\\
\mathbf{P}_{em} &  =\frac{e_{1}e_{2}}{2c^{2}}\left(  \frac{\mathbf{v}%
_{1}+\mathbf{v}_{2}}{|\mathbf{r}_{1}-\mathbf{r}_{2}|}+\frac{(\mathbf{v}%
_{1}+\mathbf{v}_{2})\cdot(\mathbf{r}_{1}-\mathbf{r}_{2})(\mathbf{r}%
_{1}-\mathbf{r}_{2})}{|\mathbf{r}_{1}-\mathbf{r}_{2}|^{3}}\right)  \label{P1}%
\end{align}
If the particles are separated along the $y$-axis in $S$, then they will be
separated along the $y$-axis in $S^{\prime}($separation perpendicular to the
velocity) and the electromagnetic energy and momentum contributions are%
\begin{align}
U_{em}^{\prime} &  =\frac{e^{2}}{l}+\frac{e^{2}V^{2}}{2c^{2}l}\text{
\ \ \ particle separation perpendicular to velocity}\label{U2}\\
\mathbf{P}_{em}^{\prime} &  \mathbf{=}\mathbf{-}\frac{e^{2}}{lc^{2}}%
\mathbf{V}\label{P2}%
\end{align}
since $\mathbf{v}_{1}\cdot(\mathbf{r}_{1}-\mathbf{r}_{2})\mathbf{v}_{2}%
\cdot(\mathbf{r}_{1}-\mathbf{r}_{2})=[(-\widehat{i}V)\cdot\widehat
{j}l][(-\widehat{i}V)\cdot\widehat{j}l]=0$. \ These terms in Eqs. (\ref{U2})
and (\ref{P2}) agree with the terms in the expected 4-vector form in Eqs.
(\ref{Uex}) and (\ref{Pex}). \ On the other hand, if the charges are separated
along the $x$-axis (separation parallel to the velocity), then $\mathbf{v}%
_{1}\cdot(\mathbf{r}_{1}-\mathbf{r}_{2})\mathbf{v}_{2}\cdot(\mathbf{r}%
_{1}-\mathbf{r}_{2})=V^{2}l^{2}$ and the electromagnetic contributions to the
energy and momentum are doubled becoming%
\begin{align}
U_{em}^{\prime} &  =\frac{e^{2}}{l}+\frac{e^{2}V^{2}}{c^{2}l}\text{
\ \ \ \ particle separation parallel to velocity}\label{U3}\\
\mathbf{P}_{em}^{\prime} &  =-\frac{2e^{2}}{lc^{2}}\mathbf{V}\label{P3}%
\end{align}
which disagrees with the naive 4-vector form in Eqs. (\ref{Uex}) and
(\ref{Pex}) by a factor of 2 in the $1/c^{2}$ terms. \ For the spherical
classical model of the the electron, the errant factor is the famous 4/3 for
the momentum; the 4/3 is intermediate between the factors of 1 and 2 found for
the point charge examples above and corresponds to averaging over the relative
orientations for the spherical charge distribution.\cite{def}

\subsubsection{Validity of Relativistic Conservation Laws}

We should note that this failure of the naive 4-vector energy-momentum
transformation is fully consistent with the relativistic conservation laws.
\ In the presence of external forces, we must use the relativistic
conservation laws given above in Eqs. (\ref{1})-(\ref{4}). \ Since the
external forces $\mathbf{F}_{1}$ and $\mathbf{F}_{2}$ are equal in magnitude
and opposite in direction, there is no net energy or linear momentum
introduced by the external forces. \ The angular momentum situation will be
discussed below. \ Since $\mathbf{F}_{A}=-\mathbf{F}_{B},$ the fourth
conservation law (\ref{4}) in frame $S^{\prime}$ here gives%
\begin{equation}
\mathbf{F}_{1}^{\prime}\cdot(-\mathbf{V)(r}_{1}^{\prime}-\mathbf{r}%
_{2}^{\prime})/c^{2}=U^{\prime}(-\mathbf{V)/}c^{2}\mathbf{-P}^{\prime
}\label{f1}%
\end{equation}
If the charges are separated along the $y$-axis perpendicular to the velocity,
then $\mathbf{F}_{1}^{\prime}$ is perpendicular to $\mathbf{V}$ and the
left-hand side of Eq. (\ref{f1}) vanishes giving $\mathbf{P}^{\prime}%
=(U/c^{2})(-\mathbf{V})$ which indeed holds in connection with Eqs.
(\ref{U0}), (\ref{P0}), (\ref{U2}) and (\ref{P2}). \ If the charges are
separated along the $x$-direction parallel to the velocity, then the left-hand
side of Eq. (\ref{f1}) can be found from the force transformation Eq.
(\ref{F1}) (and the Lorentz contraction in the direction of motion)
$\mathbf{F}_{A}^{\prime}\cdot(-\mathbf{V)(r}_{A}^{\prime}-\mathbf{r}%
_{B}^{\prime})/c^{2}=\widehat{i}(e^{2}/l^{2})(Vl/c^{2})=\widehat{i}%
(e^{2}/l)(V/c^{2})$ which is exactly the additional term needed to connects
Eqs. (\ref{U0}), (\ref{P0}), (\ref{U3}) and (\ref{P3}) through Eq.
(\ref{f1}).. \ 

\subsubsection{Particles Approaching Each other Along the $x$-Axis}

A physical understanding of the failure of the naive energy-momentum 4-vector
transformation when external forces are present can be obtained by considering
the formation of the system when the particles are sent towards each other
from spatial infinity, and then the external forces are applied as the
particles reach equilibrium positions.

First we consider two point particles, both of charge $e$ and mass $m$
approaching the origin symmetrically along the $x$-axis in the inertial frame
$S$. \ The left-hand particle has velocity $\mathbf{v}_{1}=\widehat{i}v_{0}$
and the right-hand particle has velocity $\mathbf{v}_{2}=-\widehat{i}v_{0}$ so
that the system linear momentum and angular momentum about the origin are both
zero%
\begin{equation}
\mathbf{P}_{0}=0,\text{ \ \ \ \ }\mathbf{L}_{0}=0\label{f2}%
\end{equation}
\ The particles start out at spatial infinity with speed $v_{0},$ and then the
particles slow down as they repel each other and the kinetic energy is
converted into electric potential energy. \ In the order of approximation
$v^{2}/c^{2}$ of the Darwin Lagrangian, there is no radiation energy loss due
to the acceleration of the charges. \ At time $t=0$ in the $S$ inertial frame,
both the particles come to rest at a separation $l,$ $\mathbf{r}%
_{1}(0)=-\widehat{i}l/2=-$ $\mathbf{r}_{2}(0),$ where the electrostatic
potential energy equals the initial kinetic energy $KE_{0}=U_{0}-2mc^{2},$%
\begin{equation}
KE_{0}=2\left[  \frac{1}{2}m\left(  1+\frac{3}{4}\frac{v_{0}^{2}}{c^{2}%
}\right)  v_{0}^{2}\right]  =\frac{e^{2}}{l}\label{f3}%
\end{equation}
When the particles come to rest, external forces $\mathbf{F}_{1}=\widehat
{i}e^{2}/l^{2}=-$ $\mathbf{F}_{2}$ are applied simultaneously to the left- and
right-hand particles respectively, so as to balance the electrostatic
repulsion and to keep the particles at rest. \ Since these particles are
applied at the same time in the $S$ frame when the particles are at rest, the
forces introduce neither energy nor linear momentum nor angular momentum into
the system of the two point particles. \ Thus in the $S$ frame, after the
external forces are applied, the final energy $U_{F}$ equals the initial
energy $U_{0},$ $U_{F}=U_{0},$ the final momentum equals the initial momentum,
$\mathbf{P}_{F}=\mathbf{P}_{0}=0,$ and the final angular momentum equals the
initial angular momentum, $\mathbf{L}_{F}=\mathbf{L}_{0}=0.$

However, now consider this same situation from the point of view of an
observer in an inertial frame $S^{\prime}$ moving with velocity $\mathbf{V=}%
\widehat{i}V$ relative to the inertial frame $S.$ \ When the particles are far
apart, the velocities of the particles in the $S^{\prime}$ frame are along the
$x^{\prime}$-axis%
\begin{equation}
v_{1}^{\prime}=\frac{v_{0}-V}{1-v_{0}V/c^{2}}\approx(v_{0}-V)(1+v_{0}%
V/c^{2})\label{f4}%
\end{equation}%
\begin{equation}
v_{2}^{\prime}=\frac{-v_{0}-V}{1+v_{0}V/c^{2}}\approx(-v_{0}-V)(1-v_{0}%
V/c^{2})\label{f5}%
\end{equation}
through order $v^{2}/c^{2}.$ \ Then in the $S^{\prime}$ frame when the
particles are far apart, the total energy and momentum are found to be%
\begin{align}
U_{0}^{\prime} &  =2mc^{2}+\frac{1}{2}m\left(  \mathbf{v}_{1}^{\prime2}%
+\frac{3}{4}\frac{(\mathbf{v}_{1}^{\prime2})^{2}}{c^{2}}\right)  +\frac{1}%
{2}m\left(  \mathbf{v}_{2}^{\prime2}+\frac{3}{4}\frac{(\mathbf{v}_{2}%
^{\prime2})^{2}}{c^{2}}\right)  \nonumber\\
&  =2mc^{2}+\frac{1}{2}m\left\{  \left[  (v_{0}-V)\left(  1+\frac{v_{0}%
V}{c^{2}}\right)  \right]  ^{2}+\frac{3}{4c^{2}}\left[  (v_{0}-V)\left(
1+\frac{v_{0}V}{c^{2}}\right)  \right]  ^{4}\right\}  \nonumber\\
+ &  \frac{1}{2}m\left\{  \left[  (-v_{0}-V)\left(  1-\frac{v_{0}V}{c^{2}%
}\right)  \right]  ^{2}+\frac{3}{4c^{2}}\left[  (-v_{0}-V)\left(
1-\frac{v_{0}V}{c^{2}}\right)  \right]  ^{4}\right\}  \nonumber\\
&  =\left(  1+\frac{1}{2}\frac{V^{2}}{c^{2}}+\frac{3V^{4}}{8c^{4}}\right)
2\left[  mc^{2}+\frac{1}{2}m\left(  \mathbf{v}_{0}^{2}+\frac{3}{4}%
\frac{(\mathbf{v}_{0}^{2})^{2}}{c^{2}}\right)  \right]  \approx\gamma_{V}%
U_{0}-\gamma_{V}VP_{0}\nonumber\\
&  =\left(  1+\frac{1}{2}\frac{V^{2}}{c^{2}}+\frac{3V^{4}}{8c^{4}}\right)
\left(  2mc^{2}+\frac{e^{2}}{l}\right)  \label{f7}%
\end{align}
and
\begin{align}
\mathbf{P}^{\prime} &  =m\left(  1+\frac{v_{A}^{\prime2}}{2c^{2}}\right)
\mathbf{v}_{A}+m\left(  1+\frac{v_{B}^{2}}{2c^{2}}\right)  \mathbf{v}%
_{B}\nonumber\\
&  =\widehat{i}m\left\{  1+\frac{1}{2c^{2}}\left[  (v_{0}-V)\left(
1+\frac{v_{0}V}{c^{2}}\right)  \right]  ^{2}\right\}  (v_{0}-V)\left(
1+\frac{v_{0}V}{c^{2}}\right)  \nonumber\\
&  +\widehat{i}m\left\{  1+\frac{1}{2c^{2}}\left[  (-v_{0}-V)\left(
1-\frac{v_{0}V}{c^{2}}\right)  \right]  ^{2}\right\}  (-v_{0}-V)\left(
1-\frac{v_{0}V}{c^{2}}\right)  \nonumber\\
&  =-\left(  1+\frac{1}{2}\frac{V^{2}}{c^{2}}\right)  \mathbf{V}\frac{U_{0}%
}{c^{2}}\approx\widehat{i}\left(  \gamma_{V}P_{0}-\frac{\gamma_{V}VU_{0}%
}{c^{2}}\right)  \nonumber\\
&  =-2m\left(  1+\frac{1}{2}\frac{V^{2}}{c^{2}}\right)  \mathbf{V-}\frac
{e^{2}}{lc^{2}}\mathbf{V}\label{f8}%
\end{align}
where in the last lines we have used Eq. (\ref{f3}) connecting the initial
kinetic energy and the final potential energy is $S$. \ Thus when the
particles are far apart, the total energy and momentum of the system transform
as a Lorentz four-vector between the $S$ and the $S^{\prime}$ inertial frames. \ 

In the absence of external forces, the energy and momentum are constant in
each inertial frame. \ Therefore the four-vector transformation character
between the frames $S$ and $S^{\prime}$ is retained until the application of
the external forces. \ However, when the external forces are applied
simultaneously in the $S$ frame, they are not applied simultaneously in the
$S^{\prime}$ frame. \ Rather the force $\mathbf{F}_{1}^{\prime}$ is applied at
time $t_{1}^{\prime}=\gamma_{V}[0+Vl/(2c^{2})]\approx Vl/(2c^{2})$ while the
force $\mathbf{F}_{2}^{\prime}$ is applied at time $t_{2}^{\prime}=\gamma
_{V}[0-Vl/(2c^{2})]\approx-Vl/(2c^{2}),$ where we have used the Lorentz
transformations through order $v^{2}/c^{2}.$ \ Thus in the $S^{\prime}$ frame
the application of the forces is not simultaneous but rather the right-hand
force is applied first and acts alone during a time interval $\delta
t^{\prime}=t_{2}^{\prime}-t_{1}^{\prime}=Vl/c^{2}.$ \ Thus during the time
interval $\delta t^{\prime}$ there is a net energy $\delta U^{\prime}$
delivered by the unbalanced force $\mathbf{F}_{2}$
\begin{equation}
\delta U^{\prime}=%
{\textstyle\int}
\mathbf{F}_{2}\cdot\mathbf{v}_{2}dt^{\prime}=\frac{e^{2}}{l^{2}}V\frac
{Vl}{c^{2}}=\frac{e^{2}V^{2}}{lc^{2}}\label{f9}%
\end{equation}
and a net momentum
\begin{equation}
\delta\mathbf{P}^{\prime}=%
{\textstyle\int}
\mathbf{F}_{2}dt^{\prime}=-\widehat{i}\frac{e^{2}}{l^{2}}\frac{Vl}{c^{2}%
}=-\widehat{i}\frac{e^{2}V}{lc^{2}}\label{f10}%
\end{equation}
Thus after both external forces are applied, the final energy $U_{F}%
^{^{\prime}}$ is given by
\begin{equation}
U_{F}^{\prime}=U_{0}^{\prime}+\delta U^{\prime}=\gamma_{V}U_{0}+\frac
{e^{2}V^{2}}{lc^{2}}=2mc^{2}\left(  1+\frac{V^{2}}{2c^{2}}+\frac{3V^{4}%
}{8c^{4}}\right)  +\left(  1+\frac{V^{2}}{2c^{2}}\right)  \frac{e^{2}}%
{l}+\frac{e^{2}V^{2}}{lc^{2}}\label{f11}%
\end{equation}
the final momentum $\mathbf{P}_{F}^{\prime}$ is given by
\begin{equation}
\mathbf{P}_{F}^{\prime}=\mathbf{P}_{0}^{\prime}+\delta\mathbf{P}^{\prime
}=\widehat{i}\left(  -\frac{\gamma_{V}VU_{0}}{c^{2}}\right)  -\widehat{i}%
\frac{e^{2}V}{lc^{2}}=-\widehat{i}\left(  1+\frac{V^{2}}{2c^{2}}\right)
\left(  2m+\frac{e^{2}}{lc^{2}}\right)  V-\widehat{i}\frac{e^{2}V}{lc^{2}%
}\label{f12}%
\end{equation}
The extra contributions $\delta U^{\prime}$ and $\delta\mathbf{P}^{\prime}$ in
Eqs. (\ref{f9}) and (\ref{f10}) correspond exactly to the changes between Eqs.
(\ref{U2}), (\ref{P2}) and Eqs. (\ref{U3}), (\ref{P3}). \ The final angular
momentum is still zero. \ Thus the system energy and momentum in the
$S^{\prime}$ frame are changed on the non-simultaneous application of the
external forces, and the energy and momentum in $S^{\prime}$ are no longer
related by four-vector Lorentz transformation to the unchanged energy and
momentum in the $S$ frame. \ Nevertheless the conservation laws connecting the
external forces to the changes of energy, of linear momentum, and of angular
momentum are valid in each inertial frame. \ This example provides a physical
understanding as to why the system energy and momentum may not transform as a
Lorentz four-vector between inertial frames when there are external forces present.

\subsubsection{Particles Approaching Each Other Along the $y$-Axis}

We can also consider the case where the two charged particles approach each
other symmetrically along the $y$-axis, with initial velocities given by
$\mathbf{v}_{1}=\widehat{j}v_{0},$ $\mathbf{v}_{2}=-\widehat{j}v_{0},$ in the
$S$ frame, while the inertial frame $S^{\prime}$ still has velocity
$\mathbf{V}=\widehat{i}V$ along the $x$-axis of the $S$ frame. \ In the $S$
frame at time $t=0$, the particles again come to rest at a separation $l,$
$\mathbf{r}_{1}(0)=-\widehat{j}l/2=-\mathbf{r}_{2}(0),$ and the external
forces $\mathbf{F}_{1}=\widehat{j}e^{2}/l^{2}=-\mathbf{F}_{2}$ are applied
simultaneously so as to maintain the particles at rest in the $S$ frame. \ In
this case, the external forces are also applied simultaneously in the
$S^{\prime}$ frame, and so do not introduce net energy or linear momentum or
angular momentum in the $S^{\prime}$ frame. \ Thus for this situation, the
energy and momentum of the system are connected as a Lorentz four-vector
between the $S$ and $S\prime$ inertial frames both before and after the
application of the external forces. \ This is in agreement with the result of
Eqs. (\ref{U2}) and (\ref{P2}).

\subsubsection{Particles Approaching Each Other Along the Line $x=y$}

Finally, we consider the two charged particles approaching each other
symmetrically along the line $x=y,$ $z=0$ in the $S$ inertial frame.
\ Initially the particles are very far apart and have velocities
$\mathbf{v}_{1}=(\widehat{i}+\widehat{j})v_{0}/2^{1/2}=-\mathbf{v}_{2}.$
\ Once again, at time $t=0,$ the particles come instantaneously to rest at
separation $l$ with $\mathbf{r}_{1}(0)=-(\widehat{i}+\widehat{j}%
)l/2^{1/2}=-\mathbf{r}_{2}(0).$ \ The external forces $\mathbf{F}%
_{1}=(\widehat{i}+\widehat{j})e^{2}/(2^{1/2}l^{2})=-\mathbf{F}_{2}$ are
applied simultaneously at $t=0,$ and so, in $S,$ the system energy, linear
momentum, and angular momentum are not changed by the simultaneous application
of the forces. \ The forces $\mathbf{F}_{1}$ and $\mathbf{F}_{2}$ are along
the 45$^{o}$-line joining the particles and so do not introduce any angular
momentum. \ However, in the $S^{\prime}$ frame, the forces are not applied
simultaneously, so that the $x$-components of the forces introduce energy and
momentum in the $S^{\prime}$ frame. \ However, in this case where the
particles are separated along the line $x=y$, there is a new aspect involving
torques and angular momentum. In this case, the forces in $S^{\prime}$ are not
colinear with the line joining the particles. \ It is this last aspect, the
non-colinearity of the forces, which we wish to emphasize at this point. 

\subsection{\ Angular Momentum and Torques in Different Inertial Frames}

\subsubsection{Two Point Charges Held Along the Line $x=y$}

Let us consider the situation of the two point charges along the line $x=y,$
$z=0$ after the application of the external forces. \ In the center-of-energy
frame, the particles are at rest and the forces are colinear with the line
separating the particles so in this frame, the angular momentum of the system
vanishes, $\mathbf{L}=0,$ and the the net torque due to external forces
vanishes. \ The conservation law for angular momentum (\ref{3}) holds
trivially in the $S$ inertial frame. \ 

The positions $\mathbf{r}_{1}^{\prime}$ and $\mathbf{r}_{2}^{\prime}$ of the
two particles in the $S^{\prime}$ frame can be found by Lorentz transformation
from the known positions $\mathbf{r}_{1}(0)=-(\widehat{i}+\widehat
{j})l/2^{1/2}=-\mathbf{r}_{2}(0)$ in the $S$ inertial frame. \ Thus we find
from $x=\gamma_{V}(x^{\prime}+Vt^{\prime})$ and $y=y^{\prime}$
\begin{equation}
\mathbf{r}_{1}^{\prime}=\widehat{i}x_{1}^{\prime}+\widehat{j}y_{1}^{\prime
}=\widehat{i}\left(  -Vt^{\prime}+\frac{x_{1}}{\gamma_{V}}\right)
+\widehat{j}y_{1}\approx\widehat{i}\left[  -Vt^{\prime}+\left(  1-\frac{V^{2}%
}{2c^{2}}\right)  \frac{-l}{2^{1/2}}\right]  +\widehat{j}\frac{-l}{2^{1/2}%
}\label{Ta}%
\end{equation}%
\begin{equation}
\mathbf{r}_{2}^{\prime}=\widehat{i}x_{2}^{\prime}+\widehat{j}y_{2}^{\prime
}=\widehat{i}\left(  -Vt^{\prime}+\frac{x_{2}}{\gamma_{V}}\right)
+\widehat{j}y_{2}\approx\widehat{i}\left[  -Vt^{\prime}+\left(  1-\frac{V^{2}%
}{2c^{2}}\right)  \frac{l}{2^{1/2}}\right]  +\widehat{j}\frac{l}{2^{1/2}%
}\label{Tb}%
\end{equation}
The external forces $\mathbf{F}_{1}^{\prime}$ and $\mathbf{F}_{2}^{\prime}$ on
the particles in the $S^{\prime}$ frame can be found by Lorentz transformation
from Eqs. (\ref{F1}) and (\ref{F2}),
\begin{equation}
\mathbf{F}_{1}^{\prime}=\frac{(\widehat{i}+\widehat{j}/\gamma)e^{2}}%
{2^{1/2}l^{2}}\approx\left[  \widehat{i}+\widehat{j}\left(  1+\frac{V^{2}%
}{2c^{2}}\right)  \right]  \frac{e^{2}}{2^{1/2}l^{2}}=-\mathbf{F}_{2}^{\prime
}\label{T1}%
\end{equation}
However, these forces are not along the line joining the particles. \ Thus in
the $S^{\prime}$ inertial frame, the external forces $\mathbf{F}_{1}^{\prime}$
and $\mathbf{F}_{2}^{\prime}$ give a net torque, whereas there is no torque in
the $S$ frame. \ Specifically, the net external torque in the $S^{\prime}$
frame is given by%
\begin{align}
&  \mathbf{r}_{A}^{\prime}\times\mathbf{F}_{A}^{\prime}+\mathbf{r}_{B}%
^{\prime}\times\mathbf{F}_{B}^{\prime}\nonumber\\
&  =\left\{  \widehat{i}\left[  -\frac{l}{2^{1/2}}\left(  1-\frac{V^{2}%
}{2c^{2}}\right)  -Vt^{\prime}\right]  +\widehat{j}\left(  \frac{-l}{2^{1/2}%
}\right)  \right\}  \times\left[  \widehat{i}+\widehat{j}\left(  1+\frac
{V^{2}}{2c^{2}}\right)  \right]  \frac{e^{2}}{2^{1/2}l^{2}}\nonumber\\
&  +\left\{  \widehat{i}\left[  \frac{l}{2^{1/2}}\left(  1-\frac{V^{2}}%
{2c^{2}}\right)  -Vt^{\prime}\right]  +\widehat{j}\left(  \frac{l}{2^{1/2}%
}\right)  \right\}  \times\left[  -\widehat{i}-\widehat{j}\left(
1+\frac{V^{2}}{2c^{2}}\right)  \right]  \frac{e^{2}}{2^{1/2}l^{2}}\nonumber\\
&  =\widehat{k}\frac{e^{2}V^{2}}{2c^{2}l}\label{T1a}%
\end{align}

At this point we consider the conservation law (\ref{3}) connecting this net
torque due to external forces to the rate of change of angular momentum in the
$S^{\prime}$ inertial frame. \ The time rate of change of the system angular
momentum follows from Eq. (\ref{e4}) as%
\begin{align}
&  \frac{d}{dt^{\prime}}[\mathbf{r}_{1}^{\prime}\times\mathbf{p}_{1}^{\prime
}+\mathbf{r}_{2}^{\prime}\times\mathbf{p}_{2}^{\prime}]\nonumber\\
&  =\mathbf{v}_{1}^{\prime}\times\mathbf{p}_{1}^{\prime}+\mathbf{v}%
_{2}^{\prime}\times\mathbf{p}_{2}^{\prime}=\mathbf{v}_{1}^{\prime}%
\times(\mathbf{p}_{1}^{\prime}+\mathbf{p}_{2}^{\prime})\nonumber\\
&  =-\widehat{i}V\times\left\{  -\widehat{i}2m\left(  1+\frac{V^{2}}{2c^{2}%
}\right)  V+2\frac{e^{2}}{2c^{2}}\left[  \frac{-\widehat{i}V}{l}%
+\frac{-\widehat{i}V\cdot(\widehat{i}+\widehat{j})l/2^{1/2}}{l^{3}}%
(\widehat{i}+\widehat{j})l/2^{1/2}\right]  \right\}  \nonumber\\
&  =\widehat{k}\frac{e^{2}V^{2}}{2c^{2}l}\label{T5}%
\end{align}
Thus indeed the conservation law $%
{\textstyle\sum_{i}}
\mathbf{r}_{i}\times\mathbf{F}_{ext\text{ }i}=d\mathcal{L}/dt$ relating the
external torque to the rate of change of angular momentum holds in both the
$S$ and the $S^{\prime}$ inertial frames, even though there is zero external
torque in one frame and non-zero torque in the other.

In Mansuripur's article, it is implied that the absence of a torque in one
inertial frame is inconsistent with the existence of a torque in a second
inertial frame.\cite{Mansuripur} \ However, precisely this situation is
involved in the Trouton-Noble situation which is considered in our example.
\ There is no inconsistency with relativity or with conservation laws for this situation.

\subsubsection{Comments on Stresses on a Rod Holding the Charges at Fixed
Positions}

In the examples above, we have considered \textit{external} forces applied to
point particles. \ In this case, the relativistic (through order $v^{2}%
/c^{2})$ analysis can be carried out in a transparent fashion. \ The original
analysis by Trouton and Noble considered charged capacitor plates (in place of
the point charges used here) which are held in place by a connecting bar.
\ Such a bar could also be introduced into the point charge analysis given in
this article.\cite{Franklin} \ In the $S$ inertial frame, the stresses in this
bar would be of nonrelativistic order to balance the electrostatic forces
between the charges. \ However, the detailed treatment of these stresses
throughout the bar does not appear to be an easy problem. \ Furthermore, the
relativistic transformation of these unknown stresses over to the $S^{\prime}$
frame does not seem trivial. \ Thus the transparent relativistic treatment
given above involving \textit{external} forces must give way to a far more
detailed and complicated treatment if the bar (with its \textit{internal}
stresses) is included as part of the system.

In any case, our simple examples do give accurate relativistic insights. \ In
particular, Marsuripur's implication that a vanishing torque in one inertial
frame is inconsistent with a non-zero torque in another inertial frame is not valid.

\section{Model for a Magnetic Moment Relativistic to Order $v^{2}/c^{2}$}

\subsection{Magnetic Moment in its Center-of-Energy Frame}

We now turn away from the torque aspect of Mansuripur's claims of
"incompatibility" over to a consideration of the magnetic moment aspect. \ To
start, we introduce a model of a magnetic moment which is transparently
relativistic to order $v^{2}/c^{2}.$

A relativistic model for a magnetic moment can be given in terms of the
uniform circular motion of two charged point particles of charges $\pm e$ and
masses $m$ and $M$ moving about the center of energy of the system. \ In an
inertial frame $S$, the first particle $m$ is located at $\mathbf{r}%
_{m}(t)=\widehat{i}r_{m}\cos(\omega t+\phi)+\widehat{j}r_{m}\sin(\omega
t+\phi)$ and the second particle $M$ is at $\mathbf{r}_{M}(t)=-\widehat
{i}r_{M}\cos(\omega t+\phi)-\widehat{j}r_{M}\sin(\omega t+\phi).$ \ In this
frame, the total linear momentum $\mathbf{P}$ of the magnetic moment system
includes both mechanical momentum and electromagnetic field momentum as given
above in Eq. (\ref{e2})%
\begin{align}
\mathbf{P}  &  =\left[  m\left(  1+\frac{r_{m}^{2}\omega^{2}}{2c^{2}}\right)
-\frac{e^{2}}{2c^{2}(r_{m}+r_{M})}\right]  \left[  -\widehat{i}r_{m}%
\sin(\omega t+\phi)+\widehat{j}r_{m}\cos(\omega t+\phi)\right] \nonumber\\
&  -\left[  M\left(  1+\frac{r_{M}^{2}\omega^{2}}{2c^{2}}\right)  -\frac
{e^{2}}{2c^{2}(r_{m}+r_{M})}\right]  \left[  -\widehat{i}r_{M}\sin(\omega
t+\phi)+\widehat{j}r_{M}\cos(\omega t+\phi)\right]  \label{M1}%
\end{align}
In this case, the electromagnetic field momentum terms involving
$\mathbf{v}_{j}\cdot(\mathbf{r}_{i}-\mathbf{r}_{j})$ vanish because the
velocity is orthogonal to the relative displacement of the particles. \ 

In the $S$ inertial frame, the linear momentum is assumed to vanish,
$\mathbf{P}=0,$ giving the condition%
\begin{equation}
\left[  m\left(  1+\frac{r_{m}^{2}\omega^{2}}{2c^{2}}\right)  -\frac{e^{2}%
}{2c^{2}(r_{m}+r_{M})}\right]  r_{m}=\left[  M\left(  1+\frac{r_{M}^{2}%
\omega^{2}}{2c^{2}}\right)  -\frac{e^{2}}{2c^{2}(r_{m}+r_{M})}\right]
r_{M}\label{M2}%
\end{equation}
In the terms of order $v^{2}/c^{2}$, we may substitute the zero-order
(nonrelativistic) conditions
\begin{equation}
mr_{m}=Mr_{M}\label{M3}%
\end{equation}
and the zero-order (nonrelativistic) equations of motion%
\begin{equation}
m\frac{v_{m}^{2}}{r_{m}}=m\omega^{2}r_{m}=\frac{e^{2}}{(r_{m}+r_{M})^{2}%
}=M\omega^{2}r_{M}=M\frac{v_{M}^{2}}{M}\label{M4}%
\end{equation}
When we replace the $v^{2}/c^{2}$ terms involving mass in Eq. (\ref{M2}) by
using the equations of motion in Eq. (\ref{M4}), we find that Eq. (\ref{M2})
simplifies to exactly the nonrelativistic condition (\ref{M3}). \ Evidently
this condition holds not only nonrelativistically but also relativistically
through order $v^{2}/c^{2}.$

The angular frequency $\omega$ is determined by the equation of motion
(\ref{e8}) for the mass $m$ in the electric and magnetic fields of the charged
particle $M$%
\begin{align}
m\gamma\frac{v^{2}}{r_{m}}  &  \approx m\left(  1+\frac{r_{m}^{2}\omega^{2}%
}{2c^{2}}\right)  r_{m}\omega^{2}\nonumber\\
&  =\frac{e^{2}}{(r_{m}+r_{M})^{2}}\left(  1+\frac{r_{M}^{2}\omega^{2}}%
{2c^{2}}\right)  -\frac{e^{2}\omega^{2}r_{M}}{c^{2}(r_{m}+r_{M})}+\frac
{e^{2}r_{m}\omega r_{M}\omega}{c^{2}(r_{m}+r_{M})^{2}} \label{M5}%
\end{align}
Once again we may use the non-relativistic equations of motion (\ref{M4}) in
the $v^{2}/c^{2}$ term involving the mass $m.$ The condition (\ref{M5}) then
simplifies to
\begin{equation}
\omega^{2}=\frac{e^{2}}{mr_{m}(r_{m}+r_{M})^{2}+e^{2}(r_{m}^{2}+r_{M}%
^{2})/(2c^{2})} \label{M6}%
\end{equation}
Since we have $mr_{m}=Mr_{M}$ from the analysis above, the expression
(\ref{M6}) for $\omega$ is invariant if we interchange $m$ and $M$, as the
expression should be. \ We can also write the condition in the symmetrical
form%
\begin{equation}
\omega^{2}=\frac{2e^{2}}{(mr_{m}+Mr_{M})(r_{m}+r_{M})^{2}+e^{2}(r_{m}%
^{2}+r_{M}^{2})/c^{2}} \label{M7}%
\end{equation}

For this magnetic moment model, the angular momentum includes both mechanical
and electromagnetic field contributions. \ From Eq. (\ref{e4}), we find that
all the angular momentum is in the direction perpendicular to the plane of the
circular motion,%
\begin{equation}
\mathbf{L}=\widehat{k}\left[  m\left(  1+\frac{r_{m}^{2}\omega^{2}}{2c^{2}%
}\right)  r_{m}^{2}\omega+M\left(  1+\frac{r_{M}^{2}\omega^{2}}{2c^{2}%
}\right)  r_{M}^{2}\omega-\frac{e^{2}}{2c^{2}}\frac{(r_{m}^{2}+r_{M}%
^{2})\omega}{(r_{m}+r_{M})^{2}}\right]  \label{M8}%
\end{equation}
Once again we may use the nonrelativistic results in Eq. (\ref{M4}) when
evaluating the terms in order $v^{2}/c^{2}$ so that the angular momentum can
be rewritten as
\begin{equation}
\mathbf{L}=\widehat{k}\omega\left[  mr_{m}^{2}+Mr_{M}^{2}-\frac{e^{2}}{2c^{2}%
}\frac{(r_{m}^{2}r_{M}+r_{M}^{2}r_{m})}{(r_{m}+r_{M})^{2}}\right]  \label{M9}%
\end{equation}

The magnetic moment for our model is usually evaluated by time averaging and
using the expression for the magnetic dipole moment of a steady-state current
distribution $\overrightarrow{\mathbf{\mu}}=[1/(2c)]%
{\textstyle\int}
d^{3}r\,\mathbf{r}\times\mathbf{J}.$ \ We can also think of averaging over an
ensemble of systems with varying phases $\phi.$ \ Thus here we obtain%
\begin{equation}
\overrightarrow{\mathbf{\mu}}=\left\langle \frac{1}{2c}%
{\textstyle\int}
d^{3}r\,\mathbf{r}\times%
{\textstyle\sum_{i}}
e_{i}\mathbf{v}_{i}\delta^{3}(\mathbf{r-r}_{i})\right\rangle =\widehat{k}%
\frac{e\omega}{2c}(r_{m}^{2}-r_{M}^{2})\label{M10}%
\end{equation}
We notice that if the masses are equal, $m=M$, then the orbital radii are
equal, $r_{m}=r_{M},$ and the magnetic moment vanishes, $\overrightarrow
{\mathbf{\mu}}=0.$ \ On the other hand, in the limit where the mass $M$ is
very large, $M>>m,$ and accordingly $r_{M}=r_{m}(m/M)$ is very small, we find
the usual connection between the magnetic moment and the nonrelativistic
angular momentum,%
\begin{equation}
\overrightarrow{\mathbf{\mu}}=\frac{e\mathbf{L}}{2mc}\label{M11}%
\end{equation}
through order $v^{2}/c^{2}.$

\subsection{Appearance of an Electric Dipole Moment in a Moving Frame}

We can also examine the appearance of our magnetic moment model when viewed
from a frame $S^{\prime}$ moving with velocity $\mathbf{V}=\widehat{i}V$
relative to $S,$ the center-of-energy frame for the magnetic moment. \ The
$x$-coordinates are related by Lorentz transformation as $x_{m}=\gamma
_{V}(x_{m}^{\prime}+Vt^{\prime})$, giving through order $1/c^{2}$%
\begin{equation}
x_{m}^{\prime}=-Vt^{\prime}+(1-V^{2}/c^{2})^{1/2}x_{m}\approx-Vt^{\prime
}+[1-V^{2}/(2c^{2})]r_{m}\cos(\omega t+\phi)\label{M12}%
\end{equation}
The time coordinates are related by Lorentz transformation as $t^{\prime
}=\gamma_{V}(t-Vx_{m}/c^{2})$ so that%
\begin{equation}
t=Vx_{m}/c^{2}+(1-V^{2}/c^{2})^{1/2}t^{\prime}=Vx_{m}/c^{2}+[1-V^{2}%
/(2c^{2})]t^{\prime}\label{M13}%
\end{equation}
Then we can simplify the expression in Eq. (\ref{M12}) through order
$v^{2}/c^{2}$ as
\begin{align}
x_{m}^{\prime} &  =-Vt^{\prime}+r_{m}\cos(\omega t+\phi)-[V^{2}r_{m}%
/(2c^{2})]r_{m}\cos(\omega t+\phi)\nonumber\\
&  =-Vt^{\prime}+r_{m}\cos\{\omega\lbrack1-V^{2}/(2c^{2})]t^{\prime}%
+\phi+\omega Vx_{m}/c^{2}\}-[V^{2}r_{m}/(2c^{2})]r_{m}\cos(\omega t^{\prime
}+\phi)\nonumber\\
&  =-Vt^{\prime}+r_{m}\cos\{\omega\lbrack1-V^{2}/(2c^{2})]t^{\prime}%
+\phi\}\cos(\omega Vx_{m}/c^{2})\nonumber\\
&  -r_{m}\sin\{\omega\lbrack1-V^{2}/(2c^{2})]t^{\prime}+\phi\}\sin(\omega
Vx_{m}/c^{2})-[V^{2}r_{m}/(2c^{2})]r_{m}\cos(\omega t^{\prime}+\phi
)\nonumber\\
&  =-Vt^{\prime}+r_{m}\cos\{\omega\lbrack1-V^{2}/(2c^{2})]t^{\prime}%
+\phi\}-r_{m}(\omega Vx_{m}/c^{2})\sin\{\omega\lbrack1-V^{2}/(2c^{2}%
)]t^{\prime}+\phi\}\nonumber\\
&  -[V^{2}r_{m}/(2c^{2})]r_{m}\cos(\omega t^{\prime}+\phi)\nonumber\\
&  =-Vt^{\prime}+r_{m}\cos\{\omega\lbrack1-V^{2}/(2c^{2})]t^{\prime}%
+\phi\}-r_{m}^{2}(\omega V/c^{2})\cos(\omega t^{\prime}+\phi)\sin(\omega
t^{\prime}+\phi)\nonumber\\
&  -[V^{2}r_{m}/(2c^{2})]r_{m}\cos(\omega t^{\prime}+\phi)\label{M14}%
\end{align}
where we have noted that for small $\delta,$ cos$\delta\approx1$ and
sin$\delta\approx\delta.$ \ The expression for $x_{M}^{\prime}$ is exactly
analogous with $r_{M}$ replacing $r_{m}$ in Eq. (\ref{M14}). \ Similarly, we
can evaluate the $y$-coordinates through order $v^{2}/c^{2}$ as%
\begin{align}
y_{m}^{\prime} &  =y_{m}=r_{m}\sin(\omega t+\phi)=r_{m}\sin\{\sigma
\lbrack1-V^{2}/(2c^{2})]t^{\prime}+\phi+\omega Vx_{m}/c^{2}\}\nonumber\\
&  =r_{m}\sin\{\omega\lbrack1-V^{2}/(2c^{2})]t^{\prime}+\phi\}\cos(\omega
Vx_{m}/c^{2})\nonumber\\
&  +r_{m}\cos\{\omega\lbrack1-V^{2}/(2c^{2})]t^{\prime}+\phi\}\sin(\omega
Vx_{m}/c^{2})\nonumber\\
&  =r_{m}\sin\{\omega\lbrack1-V^{2}/(2c^{2})]t^{\prime}+\phi\}+r_{m}(\omega
Vx_{m}/c^{2})\cos\{\omega\lbrack1-V^{2}/(2c^{2})]t^{\prime}+\phi\}\nonumber\\
&  =r_{m}\sin\{\omega\lbrack1-V^{2}/(2c^{2})]t^{\prime}+\phi\}+(r_{m}%
^{2}\omega V/c^{2})\cos^{2}(\omega t^{\prime}+\phi)\label{M15}%
\end{align}
\ Also, the expression for $y_{M}$ is analogous with $r_{M}$ replacing $r_{m}$
in Eq. (\ref{M15}). \ If we (ensemble) average the expressions (\ref{M14}) and
\ref{M15}) over the phase angle $\phi$ at a single time $t^{\prime},$ we find
that $<x_{m}>=-Vt^{\prime}$ while $<y_{m}>=[r_{m}^{2}\omega V/(2c^{2})].$
\ Thus in the inertial frame $S^{\prime},$ our model for a magnetic moment has
an electric dipole moment $\overrightarrow{\mathfrak{p}}$ perpendicular to the
direction of motion of the center of energy of the system
\begin{equation}
\overrightarrow{\mathfrak{p}}\mathbf{=}\widehat{j}e(r_{m}^{2}-r_{M}^{2})\omega
V/(2c^{2})=(-\mathbf{V/}c)\times\overrightarrow{\mathbf{\mu}}\label{M16}%
\end{equation}
where $-\mathbf{V}$ is the velocity of the center of energy of the magnetic
dipole in the $S^{\prime}$ frame. \ Thus our magnetic moment model has on
average an electric dipole moment of order $v^{2}/c^{2}$ as seen in the
$S^{\prime}$ inertial frame, although the charge distribution in the $S$
inertial frame has no average electric dipole moment. \ 

\section{Comments on the Interaction of a Magnetic Dipole and Electric Charge}

The examples above illustrate some relativistic aspects (through order
$v^{2}/c^{2})$ of very simple point-charge systems. \ One system which aroses
great controversy involves a magnetic moment and a point charge. \ It is the
system invoked in the recent work by Mansuripur. \ The interaction of a point
charge and a magnetic moment seems quite complicated. \ It is evident that if
a point charge were to be introduced some distance away from our relativistic
(to order $v^{2}/c^{2})$ magnetic dipole model, the charges of the dipole
would not continue in circular motion. \ Rather there would be nonrelativistic
(zero-order in $v/c)$ interactions which would alter the magnetic moment.
\ The behavior of our model (in the limit $M>>m)~$has been discussed by
Solem\cite{Solem} in his article "The Strange Polarization of the Classical
Atom." \ The circular orbits become elliptical and lead to fascinating
unexpected forces back on the distant point charge.\cite{b2006} \ The magnetic
moment model discussed here seems to have provided the basis for the only
relativistically accurate account of a magnetic moment and a point
charge.\cite{b2006} \ Of course, the physics literature is full of accounts
claiming to describe the interaction of a magnetic moment and a point charge,
and these are listed in Griffiths' resource letter.\cite{RLetter} \ These
accounts included counter rotating disks carrying charges\cite{Shockley} or
tubes carrying charges\cite{Vaidman} or rotating rigid cylinders carrying
charges\cite{Peshkin}. \ What is common to all these accounts is the
assumption that the path taken by the charges which provide the magnetic
moment is rigid and unchanged by the introduction of the external charge or
external electric field. \ An example of exactly this fixed-path point of view
is given  in a fine undergraduate text book on electromagnetism where the
example claims to calculate the "hidden momentum" associated with a magnetic
moment.\cite{Gexample} \ In the example, the "hidden momentum" or order
$v^{2}/c^{2}$ is calculated without ever discussing the role played by the
forces which hold the moving charges in the prescribed path. \ \ All these
fixed-path accounts\cite{Glist} are suspect because there is no discussion of
the (large) nonrelativistic stresses which must be present to maintain the
fixed path in which the (small) relativistic corrections appear for the moving
charges.\cite{excuse} \ The large nonrelativistic stresses may produce small
relativistic effects which are totally different from the small mechanical
relativistic effects calculated from the fixed-path point of view. \ Indeed
exactly this situation appears in the accurate relativistic
calculations\cite{b2006} based upon the point charge magnetic-moment model
used in the present article. \ The nonrelativistic electromagnetic behavior of
the magnetic moment leads to surprising relativistic forces back on the
distant point charge\cite{b2006} which are denied in the mainstream physics
literature and are exactly of the sort to account for the Aharonov-Bohm phase
shift as arising from classical electromagnetic effects.

\section{Closing Remarks}

The present article was provoked by Mansuripur's recent article claiming that
the Lorentz force law is incompatible with relativity.\cite{other} \ His
article is the latest entry in a set of perennial problems mentioned in
Griffiths' valuable resource letter on Electromagnetic Momentum.\cite{RLetter}
\ Apparently there are a number of aspects of classical electromagnetism which
are sufficiently unfamiliar that physicists rediscover them and then comment
on what seems an unusual situation. \ One of the oldest conundrums is that of
the "4/3 problem" for the classical model of the electron, involving the
failure of 4-vector Lorentz transformation properties for energy-momentum when
external forces are present. \ Another is the controversy between the Abraham
and the Minkowski tensors for momentum carried by materials. \ The
Trouton-Noble situation involving inertial-frame-dependent torques is another
perennial problem. \ Finally, the interaction of a point charge and a magnetic
moment has puzzled physicists for years, and has generated controversy
involving "hidden momentum" and in connection with the Aharonov-Bohm and
Aharonov-Casher effects. \ Mansuripur's article has comments relevant to the
last two controversies on this list. \ In the present article, we have tried
to discuss some very simple point-charge systems from a relativistic point of
view through order $v^{2}/c^{2}$ in order to provide a basic physical
understanding of what is involved in some of the controversies. \ Both the
"4/3 problem" and the Trouton-Noble situations are well, though perhaps not
widely understood; they are clearly illustrated by our simple examples. \ The
Abraham-Minkowski controversy continues to receive attention in the research
literature but rarely in the teaching literature; it is not treated in the
present article. \ The interaction of a point charge and a magnetic moment has
been discussed repeatedly in both the teaching and the research literature
with many references to "hidden momentum." \ However, we have suggested in
this article that the behavior of a magnetic moment in an electric field due
to external charges is still not properly appreciated, not even at a
nonrelativistic level.


\begin{thebibliography}{99}                                                                                               %


\bibitem {Griffiths}D. J. Griffiths, \textit{Introduction to Electrodynamics}
3rd edn (Prentice-Hall, Upper Saddle River,NJ 1999).

\bibitem {Jackson}J. D. Jackson, \textit{Classical Electrodynamics} 3rd edn
(Wiley, New York 1999).

\bibitem {Mansuripur}M. Mansuripur, "Trouble with the Lorentz law of force:
Incompatibility with special relativity and momentum conservation," Phys. Rev.
Lett. \textbf{108}, 023807 (2012)

\bibitem {Science}A. Cho, "Textbook electrodynamics may contradict
relativity," Science \textbf{336}, 404 (2012).

\bibitem {TN}F. T. Trouton and H. R. Noble, Phil. Trans. \textbf{A202}, 165-
(1903); Proc. Roy. Soc. \textbf{72}, 132- (1903).

\bibitem {TNarticles}See, for example, T. Ivezic, Am. J. Phys. \textbf{72},
1484 (2004); J. D. Jackson, Am. J. Phys. \textbf{72}, 1484 (2004); S. A.
Teukolsky, Am. J. Phys. \textbf{64} 1104 (1996).

\bibitem {Shockley}W. Shockley and R. P. James, "'Try simplest cases'
discovery of 'hidden momentum forces on 'magnetic currents'," Phys. Rev. Lett.
\textbf{18}, 876-879 (1967).

\bibitem {AB}Y. Aharonov and D. Bohm, "Significance of electromagnetic
potentials in quantum theory," Phys. Rev. \textbf{115}, 485-491 (1959).

\bibitem {AC}Y. Aharonov and A. Casher, "Topological quantum effects for
neutral particles," Phys. Rev. Lett. \textbf{53}, 319-321 (1984).

\bibitem {RLetter}D. J. Griffiths, Resource Letter EM-1: Electromagnetic
Momentum," \textbf{80}, 7-18 (2012).

\bibitem {43prob}See the list of reference in Griffiths' resource letter (ref.
10) under the heading "D. Momentum and Mass."

\bibitem {b44}T. H. Boyer, "Interaction of a Point Charge and a Magnet:
Comments on "Hidden Mechanical Momentum Due to Hidden Nonelectromagnetic
Forces'," arXiv:0708.3367v1 (2007).

\bibitem {CofE}T. H. Boyer, "Illustrations of the relativistic conservation
law for the center of energy," Am. J. Phys. \textbf{73}, 953-961 (2005). \ The
generator of Lorentz transformations is the system energy times the center of energy.

\bibitem {Implications}T. H. Boyer, "Illustrating some implications of the
conservation laws in relativistic mechanics," Am. J. Phys. \textbf{77},
562-569 (2009).

\bibitem {Darwin}See ref. 2, pp. 596-598. C. G. Darwin was the grandson of the
famous Darwin of evolution.

\bibitem {assign}The fields are given by L. Page and N. I. Adams,
\textquotedblleft Action and reaction between moving charges," Am. J. Phys.
\textbf{13}, 141--147 (1945). \ I sometimes assign undergraduate projects
which combine what students have learned in both mechanics and
electromagnetism by asking students to prove that Maxwell's equations hold
through order $v^{2}/c^{2}$ for the fields in Eqs. (\ref{e9}) and (\ref{e10}),
and also to evaluate the Poisson brackets and to show that various
conservation laws hold for the Darwin Lagrangian in Eq. (\ref{e1}).

\bibitem {b1985a}Aspects of this example were discussed earlier, T. H. Boyer,
"Lorentz-transformation properties for energy and momentum in electromagnetic
systems," Am. J. Phys. \textbf{53}, 167-171 (1985).

\bibitem {def}T. H. Boyer, "Classical model of the electron and the definition
of electromagnetic field momentum," Phys. Rev. D \textbf{25}, 3246-3250 (1982).

\bibitem {Franklin}Recently J. Franklin mentioned a "string" holding two like
charges in place, "The lack of rotation in the Trouton-Noble experiment"
arXiv:physics/0603110v3, 1-8 (2006). \ However, a "string" may bring to mind
properties which contradict the requirements of special relativity. \ 

\bibitem {Solem}J. C. Solem, "The strange polarization of the classical atom,"
Am. J. Phys. \textbf{55}, 906-909 (1987).

\bibitem {b2006}T. H. Boyer, "Darwin-Lagrangian analysis for the interaction
of a point charge and a magnet: Considerations related to the controversy
regarding the Aharonov-Bohm and Aharonov-Casher phase shifts," J. Phys.
A:Math. Gen. \textbf{39}, 3455-3477 (2006).

\bibitem {Vaidman}L. Vaidman, "Torque and force on a magnetic dipole," Am. J.
Phys. \textbf{58}, 978-983 (1990).

\bibitem {Peshkin}M. Peshkin, I Talmi, and L. J. Tassie, "the quantum
mechanical effects of magnetic fields confined to inaccessible regions," Ann.
Phys., NY \textbf{12}, 426-435 (1961), especially section 5.

\bibitem {Gexample}See ref. 1, p. 520.

\bibitem {Glist}See the list of references in Griffiths' resource letter (ref.
10) under the heading "C. Hidden Momentum."

\bibitem {excuse}Indeed, some authors seem to feel that simply making
reference to "hidden momentum" excuses them from properly accounting for the
flow of energy and momentum in an electromagnetic system.

\bibitem {other}Replies to Mansuripur's assertions have been given also by K.
T. McDonald, "Mansuripur's Paradox,"
http://www.physics.princeton.edu/\symbol{126}mcdonald/examples/mansuripur.pdf
(2012); and by D. A. T. Vanzella, "Comment on 'Trouble with the Lorentz law of
force: Incompatibility with special relativity and momentum conservation,"
arXiv:1205.1502v1 (2012).
\end{thebibliography}
\end{document}